\documentclass[10pt, aps, twocolumn]{revtex4}
\usepackage{graphicx}
\def\vector#1{\mbox{\boldmath $#1$}}

\begin{document}

\title{A model of magnetic friction obeying the Dieterich--Ruina law in the steady state }

\author{Hisato \surname{Komatsu}}
\email[Email address: ]{komatsu.hisato@nims.go.jp}

\affiliation{Research Center for Advanced Measurement and Characterization, National Institute for Materials Science, Tsukuba, Ibaraki 305-0047, Japan}

\begin{abstract}

We propose a model of magnetic friction and investigate the relation between the frictional force and the relative velocity of surfaces in the steady state. The model comprises two square lattices adjacent to each other, the upper of which is subjected to an external force, and the magnetic interaction acts as a kind of ``potential barrier'' that prevents the upper lattice from moving. We consider two surface types for the upper lattice: smooth and rough. 
The behavior of this model is classified into two domains, which we refer to as domains I and II. In domain II, the external force is dominant compared with other forces, whereas in the domain I, the the velocity of the lattice is suppressed by the magnetic interaction and obeys the Dieterich--Ruina law. This characteristic property can be observed regardless of whether the surface is smooth or rough.

\end{abstract}

\maketitle

\section{Introduction \label{Introduction} }

Friction is a very familiar phenomenon, and there have been many studies aimed at revealing its microscopic mechanisms \cite{BC06, KHKBC12}. Such studies have considered the friction due to various factors such as lattice vibration and the motion of electrons \cite{MDK94, DAK98, MK06, PBFMBMV10, KGGMRM94}. In particular, magnetic friction, which is the frictional force due to the magnetic interaction between spin variables, has been the subject of much interest \cite{WYKHBW12, CWLSJ16, LG18}, and several statistical mechanical models of magnetic friction have been proposed to date \cite{KHW08, Hucht09, AHW12, HA12, IPT11, Hilhorst11, LP16, Sugimoto19, FWN08, DD10, MBWN09, MBWN11, MAHW11}. In those studies, the relation between the frictional force and the relative surface velocity depends on the choice of model. In some cases, the relation is the Amontons--Coulomb law \cite{KHW08, Hucht09, AHW12,HA12}, in others it is the Stokes law \cite{MBWN09,MBWN11}, while in others the relation shows a crossover between these two laws \cite{MAHW11}. Those studies aimed at not only explaining the properties of magnetic friction but also revealing the microscopic mechanisms of friction, which is not restricted to the case of magnetic materials. It is therefore important to study whether the properties of other materials can be reproduced in magnetic systems. 

It is widely known that the friction between solid surfaces obey the Amontons--Coulomb law, the law that the frictional force $F$  is independent of the relative velocity $v$. However, Coulomb himself found that actual materials violate this law slightly\cite{PP15}. Empirical modification of the Amontons--Coulomb law is studied several decades ago and established as the Dieterich--Ruina law \cite{BC06, KHKBC12, Ruina83, Dieterich87, DK94, HBPCC94, Scholz98}. In the steady state, this law is expressed as 
\begin{equation}
F = A \log v + B ,
\label{DRlaw}
\end{equation}
where $A$ and $B$ are constants.
For a phenomenological derivation of the Dieterich--Ruina law, see Ref.~\cite{HBPCC94} for example. 
Such discussions suggest that a potential barrier preventing the relative surface motion is important in forming the $v$--$F$ relation of Eq.~(\ref{DRlaw}). 

In this study, we propose a new model of magnetic friction in which the structure of the spin variables prevents the relative surface motion. Here, the structure of the spin variables behaves as a potential barrier, from which we expect violation of the Amontons--Coulomb law. To consider the effect of surface roughness, we investigate lattices with either smooth or rough surfaces and compare the results. The outline of this paper is as follows. We introduce the model and how it evolves with time in Sec.~\ref{Model},  investigate the model by numerical simulation in Sec.~\ref{Simulation}, and summarize the study in Sec.~\ref{Summary}.

\section{Model \label{Model} }

\begin{figure}[hbp!]
\begin{center}
\includegraphics[width = 7.0cm]{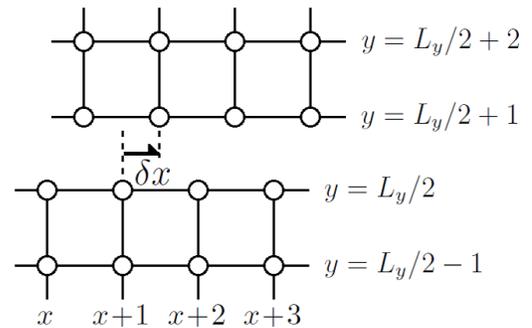}
\end{center}
\caption{Lattice arrangement considered in this study. }
\label{Lattice1}
\end{figure}

\begin{figure}[hbp!]
\begin{center}
\includegraphics[width = 7.0cm]{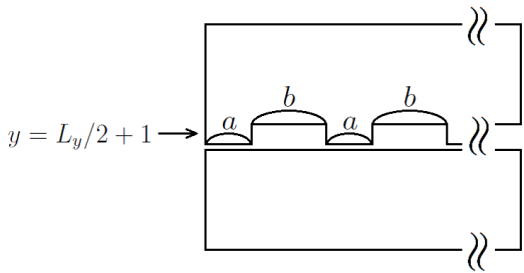}
\end{center}
\caption{Lattice shape: the upper lattice lacks $b$ lattice points after every $a$ points. A smooth surface (type-A lattice) corresponds to $b=0$. }
\label{LatticeA}
\end{figure}

We begin by preparing two $L_x \times L_y /2$ square lattices that are contiguous with each other, and we allow the upper one to move in the $x$ direction relative to the lower one as in Fig.~\ref{Lattice1}. The lattice constant is normalized as the unit length.
 Each lattice point $\vector{i} = (i_x , i_y)$ has the Ising spin $\sigma _{\vector{i} } $. To consider the effect of surface roughness, we remove some lattice points from the surface of the upper lattice so that this lattice lacks $b$ lattice points after every $a$ points at $y = L_y/2 + 1$ as in Fig.~\ref{LatticeA}. 
 In short, the lattice point $\vector{i}$ is removed when $\chi _a (i_x) = 0$, and $i_y=L_y/2+1$, where the function $\chi _{a } (n)$ is defined as
\begin{equation}
\chi _a  (n) = \left\{ 
\begin{array}{lc}
1 & \mathrm{if} \ \  n \equiv 0,1,...,a-1 \ (\mathrm{mod} (a+b)) \\
0 & \mathrm{otherwise} \\
\end{array} . \right.
\label{X1}
\end{equation}
In this study, we let $a=20$ and consider two types of upper lattice, namely, a smooth lattice (type~A) with $b=0$ and a rough lattice (type~B) with $b=20$.

\begin{figure}[hbp!]
\begin{center}
\includegraphics[width = 7.0cm]{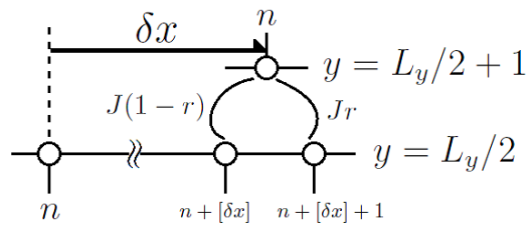}
\end{center}
\caption{Interaction between two different lattices. }
\label{Lattice2}
\end{figure}

The Hamiltonian of this system is given as
\begin{eqnarray}
{\cal H}  = & - & J \sum _{\left< \vector{i}, \vector{j} \right> } \sigma _{\vector{i}} \sigma _{\vector{j}} - J \sum _{n=1} ^{L_x}  \left\{ (1-r) \sigma _{(n+ \left[ \delta x \right] , L_y /2)}  \right. \nonumber \\ & + & \left. r\sigma _{(n+ \left[ \delta x \right] +1 , L_y /2)} \right\} \cdot \sigma _{(n, L_y /2+1)} \chi _{a} (n) \ \ ,
\label{Hamiltonian}
\end{eqnarray}
where $\delta x$ is the shift of the upper lattice, $\left[ \delta x \right]$ is the largest integer less than or equal to $\delta x$, and $r \equiv \delta x - \left[ \delta x \right]$ is the fractional part of $\delta x$. The bracket $\left< \vector{i}, \vector{j} \right> $ means the pair of adjacent spins $\vector{i}$ and $\vector{j}$ belonging to the same lattice. The second term in Eq.~(\ref{Hamiltonian}) expresses the interaction between the spins of different lattices at their boundary. Here, the spin $\sigma_{(n, L_y/2+1)}$ of the upper lattice interacts with the two nearest spins of the lower lattice, namely, $\sigma_{(n+ \left[ \delta x \right] , L_y/2)}$ and $\sigma_{(n+ \left[ \delta x \right] +1, L_y/2)}$, as in Fig.~\ref{Lattice2}. The form of Eq.~(\ref{Hamiltonian}) shows that the coupling constant between $\sigma _{(n, L_y/2+1)} $ and $\sigma _{(m, L_y/2)} $ becomes its maximum $J$ when $n+ \delta x = m$, then decreases linearly with $|n+\delta x-m|$, and finally disappears when $|n+\delta x-m| \geq 1$. The ground state of the bulk is ferromagnetic when the coupling constant $J$ is positive and antiferromagnetic when it is negative. Given that the shift of the upper lattice makes the structure of the antiferromagnetic order energetically unstable, we anticipate that the upper lattice would be difficult to move when $J$ is negative. This effect can be regarded as a kind of ``potential barrier''. In this paper, we let $J=-1$ to compare the effect of this ``potential barrier'' on the magnetic friction with that on the friction of usual solid surfaces. 

We impose the periodic boundary condition in the $x$ direction and the open boundary condition in the $y$ direction. 
The upper lattice is subjected to an external force $F$ in the $x$ direction. In the steady state, this force balances the frictional force. We let $\delta x$ obey the overdamped Langevin equation under a given temperature $T$. In this paper, we adjust the unit of the temperature so that the Boltzmann constant $k_B$ is normalized as one. Assuming that the viscous and random forces are imposed only on the boundary lattice points adjacent to the lower lattice and any elastic deformation of the lattices can be ignored, the Langevin equation for $\delta x$ is written as  
\begin{equation}
0 = - \gamma L_x ' \frac{d(\delta x)}{dt} + F - \frac{\partial {\cal H} }{\partial (\delta x)} + \sqrt{2 \gamma T L_x '} R (t) ,
\label{Langevin0}
\end{equation}
where $R$ is the white Gaussian noise fulfilling $\left< R(t) R(t') \right> = \delta (t-t')$, and $L_x '$ is the number of the lattice points of upper lattice adjacent to the lower lattice, namely
\begin{equation}
 L_x ' = \frac{a}{a+b} \cdot L_x .
\label{Lxprime}
\end{equation}
Note that all lattice points of the upper lattice move simultaneously because we ignore the elastic deformation. Using the external force per one lattice point $f \equiv F/L_x '$, we transform Eq.~(\ref{Langevin0}) as
\begin{equation}
\frac{d(\delta x)}{dt} = \frac{f}{\gamma} + \frac{1}{\gamma L_x '} \left( - \frac{\partial {\cal H} }{\partial (\delta x)} + \sqrt{2 \gamma T L_x '} R (t) \right) .
\label{Langevin1}
\end{equation}

We define the dynamics of spin variables $\sigma_i$ as updating using the Metropolis method, and we define the unit of time as one Monte Carlo step(MCS). The candidate for updating is chosen randomly at each step. For simplicity, in the actual calculation of type-B lattice, 1 MCS is defined as $L_x L_y$ steps of the updating, and we put dummy spin variable $\sigma _{\vector{i}} = 0$ which does not interact with other variables if $\vector{i}$ lacks the lattice point. The updating of $\delta x$ is done after every $\Delta t$ MCSs($=L_x L_y \Delta t$ steps) by applying the stochastic Heun method to Eq.~(\ref{Langevin1}).

In this paper, we let $\gamma = 1$, and $\Delta t = 0.01$; thus, $\delta x$ is updated after every $L_x L_y /100$ steps of updating the spin variables. The velocity $v$ of the upper lattice is defined as the change in $\delta x$ per MCS. 

Previous studies have already proposed models in which two lattices interact with each other \cite{KHW08, Hucht09, IPT11, Hilhorst11, LP16, AHW12,HA12, Sugimoto19}. However, in our model the shift of the upper lattice $\delta x$ changes according to Eq.~(\ref{Langevin1}) under a given external force $F$. This is the main difference from previous studies in which $\delta x$ has discrete values and increases according to a given constant velocity. 
These velocity-fixing models often show the saturation of energy-dissipation rate in the large-velocity limit\cite{KHW08}. As a result, the frictional force of them converges to zero in this limit. This phenomenon is observed not only in the case of the magnetic friction between two lattices, but also in the case of that between a lattice and a small tip\cite{FWN08, DD10}. This fact makes the relation between frictional force and velocity complicated and is thought to be obstructive if we try to consider the slight violation of the Amontons--Coulomb law. Our choice of model is motivated by wanting to avoid this difficulty and observe more clearly how the potential barrier prevents the motion. In the case of Eq.~(\ref{Langevin1}), in the large-$f$ limit, the first term $f/\gamma$ is thought to be dominant over the other terms of the right hand side and the velocity is thought to be proportional to $f$. Hence our model is expected not to show such complicated behavior.

\section{Simulation \label{Simulation}}

\begin{figure}[hbp!]
\begin{center}
\includegraphics[width = 8.0cm]{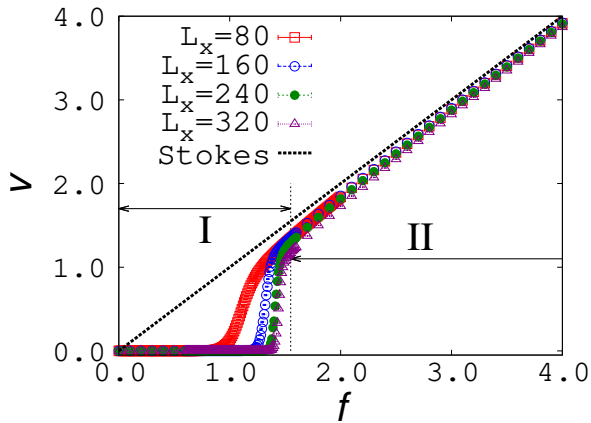} 
\end{center}
\caption{(Color online)The $f$-dependence of the velocity for type-A lattice at $T=2.0$. The symbols correspond to the data for $L_x=$80 (red squares), 160 (blue open circles), 240 (green closed circles), and 320 (purple triangles), while the heavy black dotted line corresponds to the Stokes law $v = f/\gamma$. The light dotted line drawn between domains indicate the threshold value $f_c$ which we will define later. }
\label{velo1}
\end{figure}

\begin{figure}[hbp!]
\begin{center}
\begin{minipage}{0.99\hsize}
\includegraphics[width = 8.0cm]{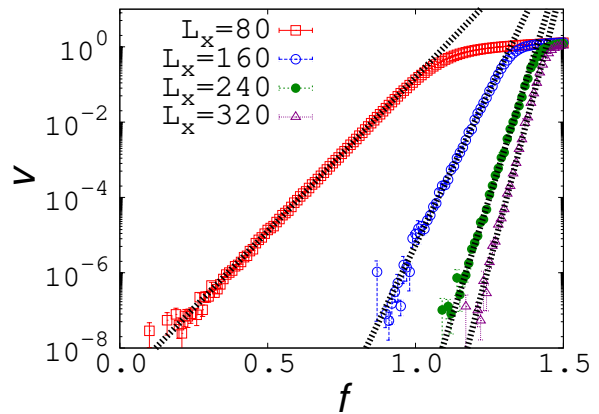} \\
\end{minipage}
\end{center}
\caption{(Color online)The $f$-dependence of the velocity for type-A lattice at $T=2.0$, plotted semi-logarithmically.  The symbols have the same meanings as in Fig.~\ref{velo1}, and the broken lines are to guide the eye. }
\label{velo2}
\end{figure}

 The numerical simulation begins from the perfect antiferromagnetic state with $\delta x = 0$, and $f$ is increased gradually. At each value of $f$, the first $2.0 \times 10^5$ MCSs is used for relaxation and next $8.0 \times 10^5$ MCSs for measurement. We take average over 96 independent trials to obtain the data with error bars. The aspect ratio of the lattice is fixed as $L_x/L_y = 8$. 
We first investigate the $f$-dependence of the velocity $v$ at $T=2.0$ of the type-A lattice, and the results are shown in Figs.\ref{velo1} and \ref{velo2}. 

\begin{figure}[hbp!]
\begin{center}
\begin{minipage}{0.99\hsize}
\includegraphics[width = 8.0cm]{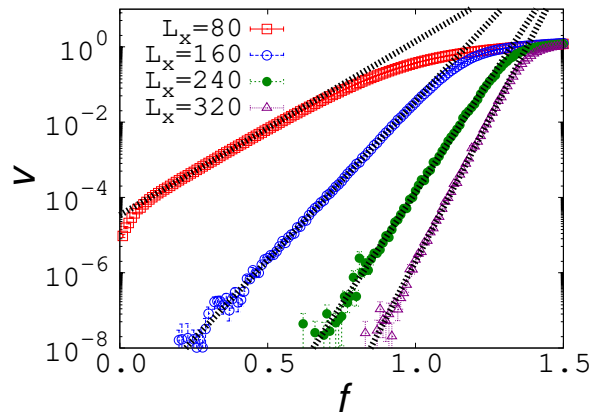} \\
\end{minipage}
\end{center}
\caption{(Color online)The $f$-dependence of the velocity for type-B lattice at $T=2.0$, plotted semi-logarithmically. The symbols have the same meanings as in Figs.~\ref{velo1} and \ref{velo2}, and the broken lines are to guide the eye. }
\label{velo2_a}
\end{figure}

In Fig.~\ref{velo1}, the $v$--$f$ curves have two domains, which we refer to as domains I and II. In domain I, the motion of the the upper lattice is suppressed by the ``potential barrier''. By contrast, in domain~II the external force dominates the other forces that appear in the right hand side of Eq.~(\ref{Langevin1}) as we anticipated in the previous section, and the $v$--$f$ curves approach the Stokes law $v = f/\gamma$ asymptotically with increasing $f$. From the semi-logarithmic graphs of the $v$--$f$ curves in Fig.~\ref{velo2}, $\log v$ seems to be a linear function of $f$ in domain~I. This relationship is consistent with the Dieterich--Ruina law in the steady state, namely Eq.~(\ref{DRlaw}). A similar investigation of the type-B lattice is shown in Fig.~\ref{velo2_a}, according to which the relation between $\log v$ and $f$ in domain~I appears to be linear even when the surface of the upper lattice is rough.

To compare these graphs with Eq.~(\ref{DRlaw}), we estimate the $v$--$F$ relation of the present model by a discussion similar to that of Ref.\cite{HBPCC94}, and rescale the graph using the estimated relation. The magnetic interaction between the two lattices causes the effective potential $U(\delta x)$ which affects the shift of upper lattice $\delta x$. Considering the effect of the external force, the potential felt by the upper lattice, $\bar{U}(\delta x)$, is expressed as
\begin{equation}
\bar{U} (\delta x) = -F \delta x + U(\delta x) .
\end{equation}
Taking the arguments of the minimum and the maximum of $\bar{U}$ as $\delta x _{\mathrm{min} } $ and $\delta x _{\mathrm{max} } $, respectively, the height of the ``potential barrier'' is expressed as 
\begin{equation}
\bar{U} (\delta x _{\mathrm{max} } ) - \bar{U} (\delta x _{\mathrm{min} } ) = -\alpha F + L' _x u_0  .
\label{barrier1}
\end{equation}
Here, $\alpha$ and $u_0$ are defined as
\begin{equation}
\alpha = \delta x _{\mathrm{max} } - \delta x _{\mathrm{min} } ,
\end{equation}
\begin{equation}
u_0 = \frac{ U (\delta x _{\mathrm{max} } ) - U (\delta x _{\mathrm{min} } ) }{L' _x}  .
\label{u0def}
\end{equation}
From the definition of $U(\delta x)$, this function is thought to be proportional to $L'_x$. We therefore infer that $u_0$ defined by Eq.~(\ref{u0def}) is independent of $L'_x$.

\begin{figure}[hbp!]
\begin{center}
\begin{minipage}{0.99\hsize}
\includegraphics[width = 4.2cm]{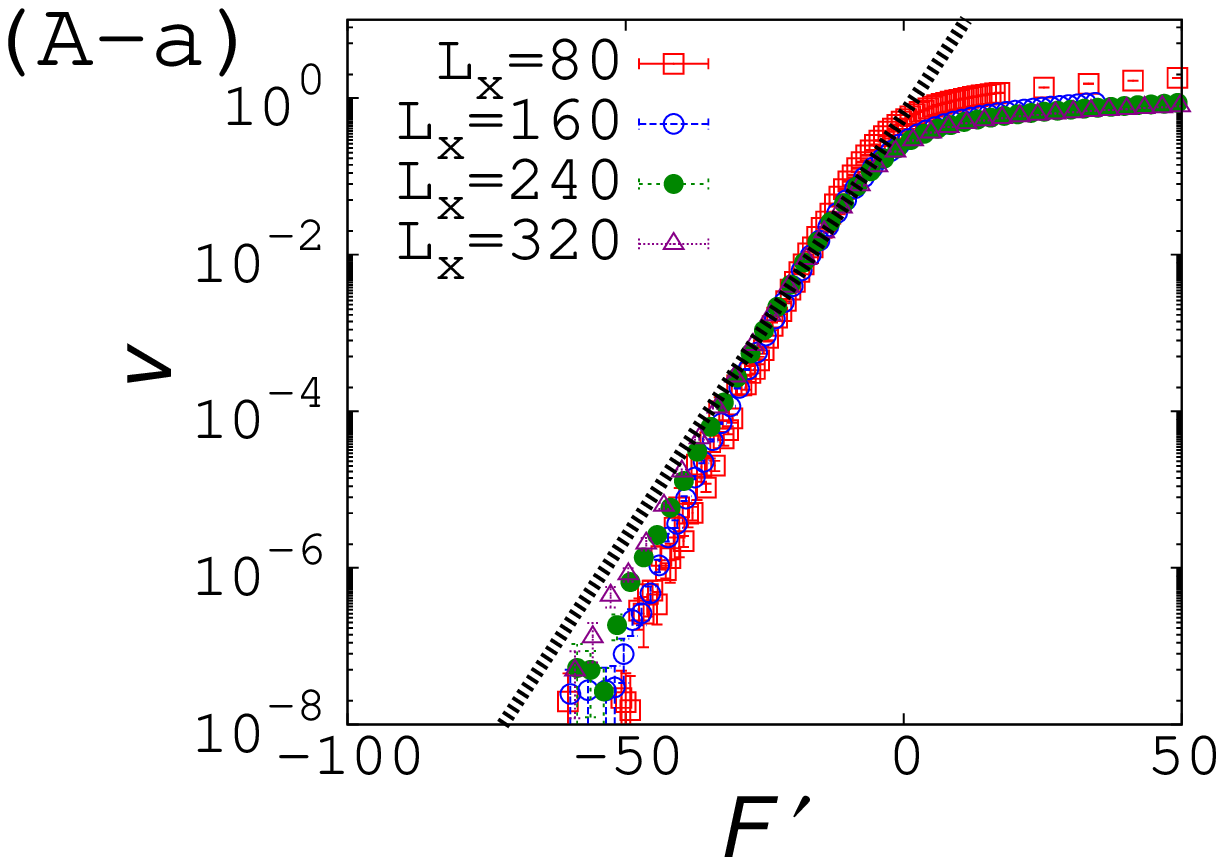} 
\includegraphics[width = 4.2cm]{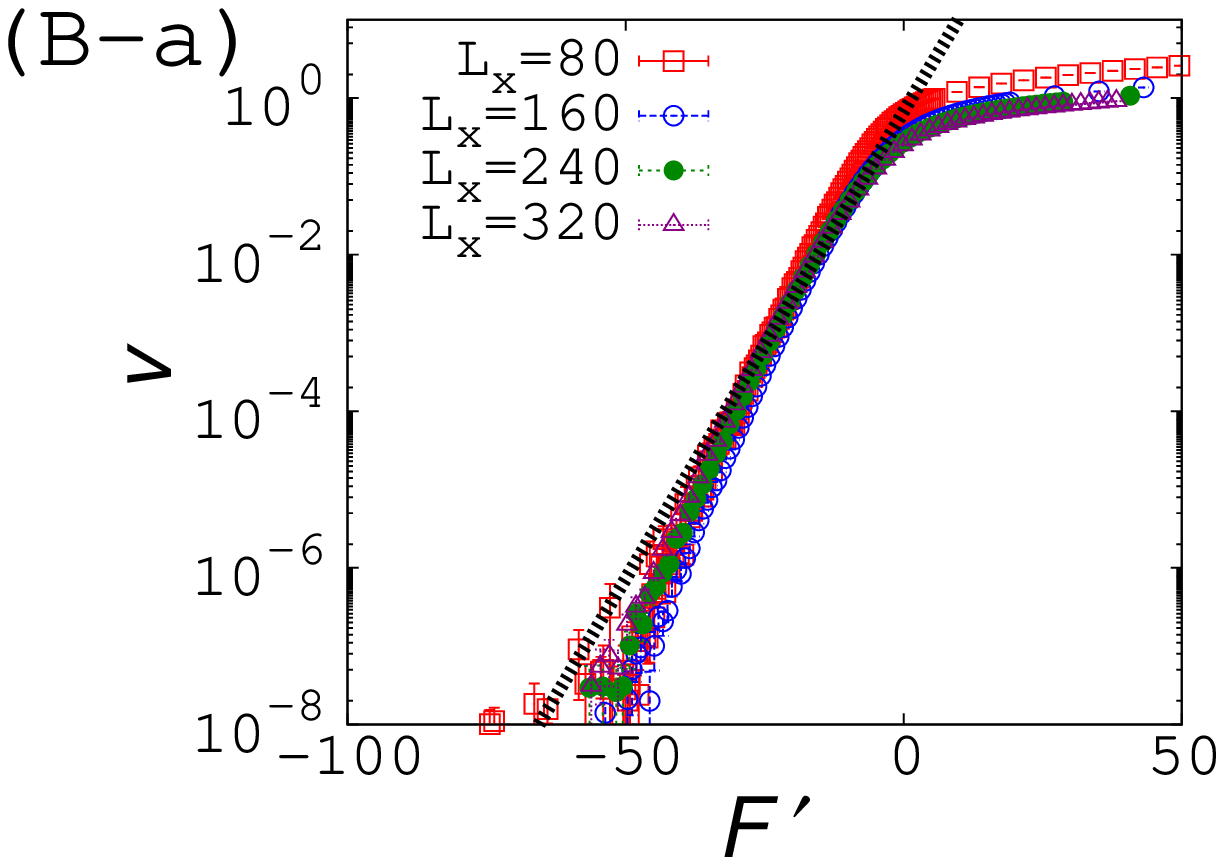} \\ 
\includegraphics[width = 4.2cm]{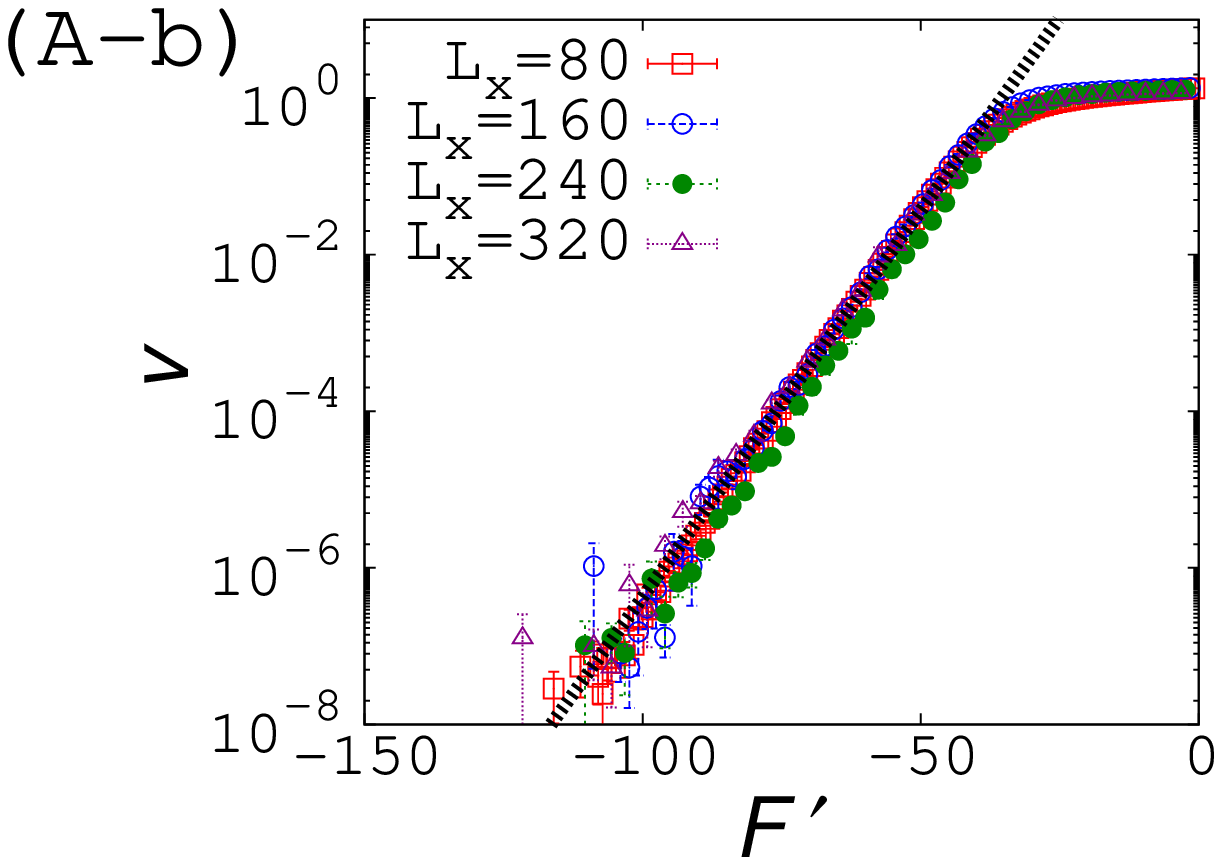} 
\includegraphics[width = 4.2cm]{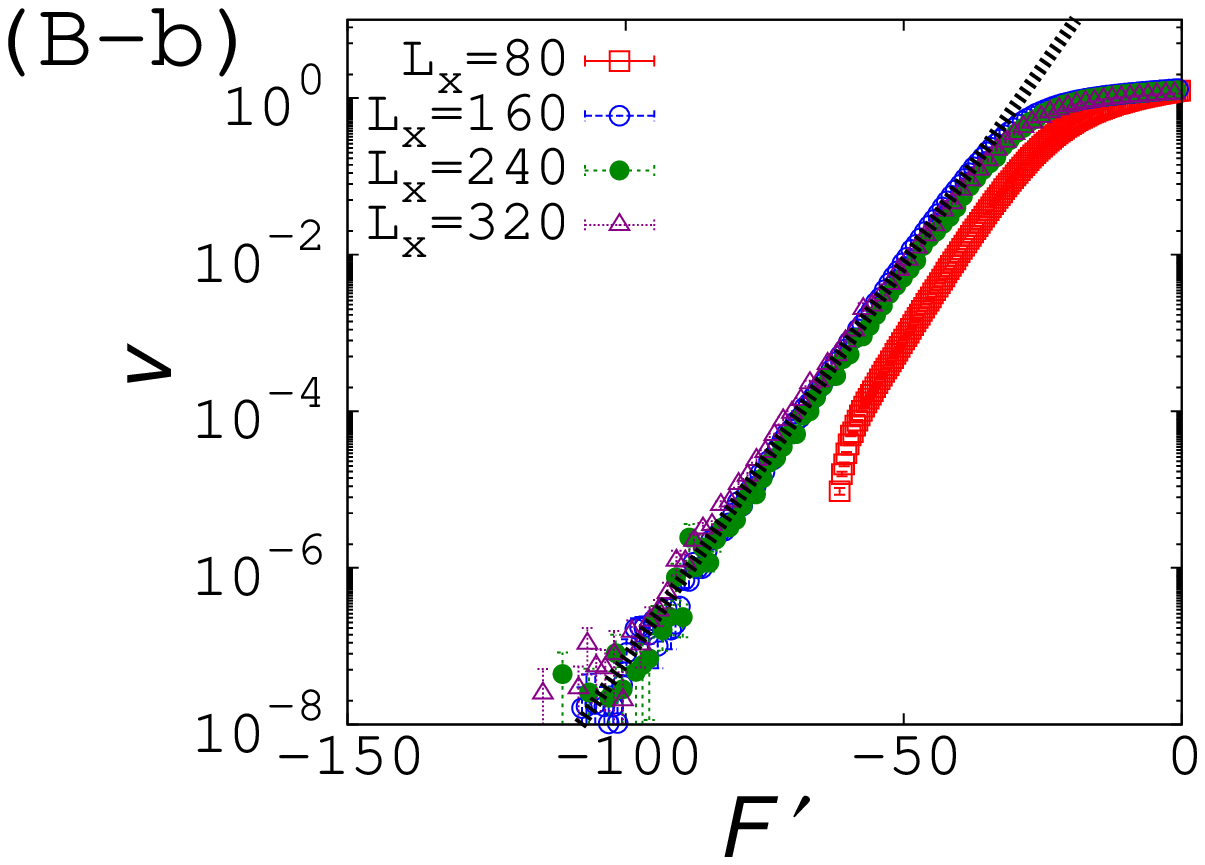} \\
\includegraphics[width = 4.2cm]{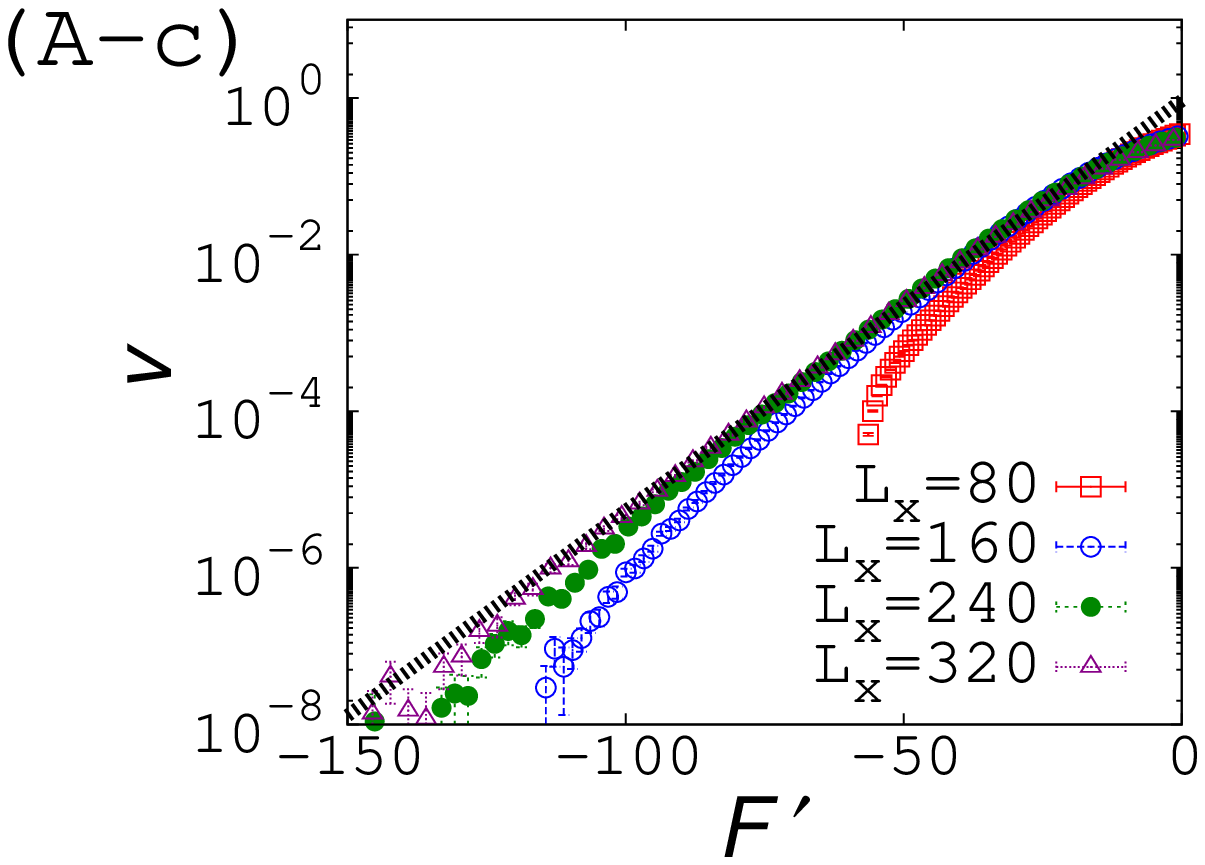} 
\includegraphics[width = 4.2cm]{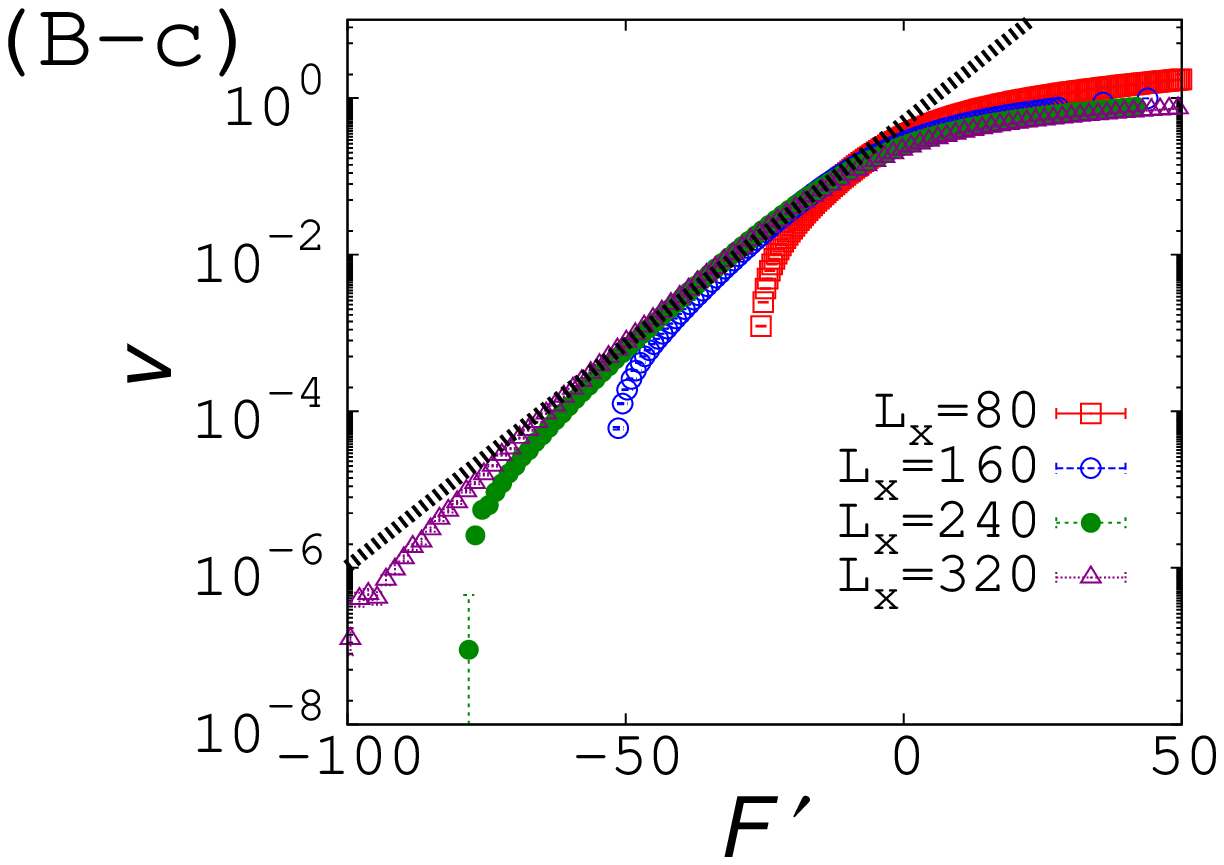} \\
\end{minipage}
\end{center}
\caption{(Color online)Rescaled $v$--$f$ relation for both lattice types at $T=1.5$ (graph (A-a) for type~A and (B-a) for type~B), 2.0 (graph (A-b) for type~A and (B-b) for type~B), and 2.5 (graph (A-c) for type~A and (B-c) for type~B) plotted semi-logarithmically. The symbols have the same meanings as in previous figures, and the broken lines are the fits of each curve in domain~II. }
\label{veloR_a}
\end{figure}

The velocity $v$ is thought to be proportional to the probability that the upper lattice acquires the sufficient energy to penetrate the ``potential barrier'' expressed by Eq.~(\ref{barrier1}), hence we have that
\begin{equation}
v = \exp \left( c -\frac{ -\alpha F + L' _x u_0 }{T} \right) , \label{vA}
\end{equation}
where, $c$ is a constant. This relation can be transformed as
\begin{equation}
\log v = \frac{ \alpha F - L' _x u_0 }{T}  +c = \alpha ' F' + c,
\label{DRlaw2}
\end{equation}
where $\alpha' = \frac{\alpha}{T} $ and 
\begin{equation}
 F' =  F - \frac{ L' _x u_0}{\alpha} = L' _x \left( f-\frac{ u_0}{\alpha} \right) . \label{Fscale}
\end{equation}
If Eq.~(\ref{DRlaw2}) is right, $v$ depends on $f$ only by the factor $F' $. Hence, to investigate whether Eq.~(\ref{DRlaw2}) holds for this model, we rescale $f$ by this factor in the $v$--$f$ relation at several temperatures for both lattice types. The constants $\alpha' , c$, and $u_0$ are determined by the least squares fitting in which we use data points that satisfy $L_x \geq 160$ and $10^{-6} \leq v \leq 10^{-1}$. The results are plotted in Fig.~\ref{veloR_a}, in which the fitted curves are drawn as the broken lines. 

\begin{figure}[hbp!]
\begin{center}
\includegraphics[width = 8.0cm]{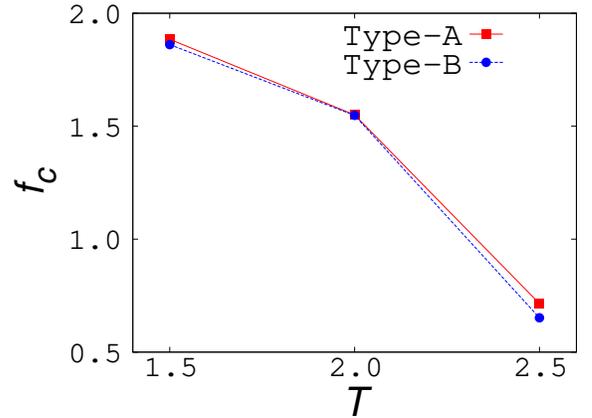} 
\end{center}
\caption{(Color online)The temperature-dependence of the threshold value $f_c$. The symbols correspond to the data for type-A(red squares) and B(blue circles) lattice. }
\label{fc1}
\end{figure}

According to these figures, the rescaled graphs for different system sizes overlap for $L_x \geq 160$. This means that Eq.~(\ref{DRlaw2}) surely holds for this model. This equation is equivalent to the Dieterich--Ruina law, namely, Eq.~(\ref{DRlaw}), with 
\begin{equation}
 A=\frac{T}{\alpha} , \ B = \frac{L' _x u_0 -cT}{\alpha} .
 \label{DR_coeff}
\end{equation}  
Consequently, our model obeys the Dieterich--Ruina law in the steady state. Judging from the fact that $f$ is scaled by $F'$ defined in Eq.~(\ref{Fscale}), $v$--$f$ curve is thought to have a discontinuous jump at a critical force $f_c \equiv \frac{u_0}{\alpha}$ in the thermodynamic limit. Hence, we regard $f_c$ as the threshold value which divide domains I and II. The temperature-dependence of $f_c$ is plotted in Fig.~\ref{fc1}. According to this figure, both of type-A and B lattices show the similar behavior of $f_c$. 

 Note that $f_c$ has the nonzero value even when $T=2.5$, the higher temperature than the equilibrium transition temperature of the two-dimensional antiferromagnetic Ising model, $T_c \simeq 2.27$. It means that the crossover or transition of the $v$--$F'$ curve from the Dieterich--Ruina law to the Stokes law is observed no matter whether the temperature is higher or lower than $T_c$. This is because the force between the two surfaces does not depend on the long-range order itself. 

Considering that the Dieterich--Ruina law is the modification of the Amontons--Coulomb law, our model seems to resemble the model of ref.\cite{MAHW11} in the point that both of them shows the change of $v$--$f$ relation from the Stokes law to the naive or modified Amontons--Coulomb law. However, the mechanisms why this change occur are different. In our model, this change is caused by whether the external force $f$ is dominant over or comparable to other forces appeared in the right hand side of Eq.~(\ref{Langevin1}).  On the other hand, that of ref.\cite{MAHW11} is caused by whether the time scale of the the relaxation of spins is faster or slower than that of the motion of the tip. Furthermore, the condition in which these two models obey the Stokes law is different from each other. The model of ref.\cite{MAHW11} obeys it when $v$ is small, unlike our model obeying it when $v$ is large.

\section{Summary \label{Summary}}

In this study, we introduced a model of magnetic friction which includes a kind of ``potential barrier'', and investigated the relation between the frictional force and the relative velocity $v$ between two surfaces in the steady state, in which the frictional force balances the external force $f$. According to the results of the numerical simulation, the upper surface moves following the Stokes law when $f$ is sufficiently strong(domain II). In contrast, the surface velocity is suppressed by the ``potential barrier'' made by the magnetic interaction, and obeys the Dieterich--Ruina law regardless of whether the surface of the upper lattice is smooth(type-A lattice) or rough(type-B lattice) when $f$ is weak(domain I). The behavior of our model is similar to the depinning transition of domain walls in the point that the energy barrier has the dominant role in the velocity $v$ of the upper lattice or the domain wall\cite{JMBSGKLJ16, DSKBJ17}. In the case of the depinning transition, as far as the external force is weaker than a certain threshold value, the velocity $v$ of the domain wall is proportional to the probability that the wall acquires the sufficient energy to penetrate the energy barrier $\Delta E$, i.e. $v \propto e^{-\Delta E /T}$. This mechanism resembles that of our discussion which derive Eq.~(\ref{vA}). 
  
However, many points remain unclear at present in order to discuss how our results relate to the friction of usual solids. For example, the original form of the Dieterich--Ruina law is not restricted to the steady state, whereas our result is. Furthermore, both of the coefficient $A$ and $B$ in Eq.~(\ref{DRlaw}) are constants for usual solids, whereas $B$ depends on the system size by $L' _x$ in Eq.~(\ref{DR_coeff}). However, we cannot identify what causes this difference by this study alone. 

The relation between the long-range order and the $v$--$f$ curve when the system contains long-range interaction also remains unclear. In the previous section, we pointed out that the qualitative behavior of the $v$--$f$ curve is not affected by whether the long-range order exists or not. However, in the case that the model contains the long-range interaction such as the dipolar interaction, the force between the two surfaces are thought to be more closely related with the long-range order.

We intend to study these problems in future work.

\section*{Acknowledgments}
Part of numerical calculations were performed on the Numerical Materials Simulator at National Institute for Materials Science. 







\begin{thebibliography}{99}


\bibitem{PP15} E.~Popova and V.~L.~Popov,
Friction \textbf{3}, 183 (2015).

\bibitem{BC06} T.~Baumberger and C.Caroli,
Adv. \ in \ Phys. \textbf{55}, 279 (2006).

\bibitem{KHKBC12} H.~Kawamura, T.~Hatano, N.~Kato, S.~Biswas, and B.~K.~Chakrabarti,
Rev. \ Mod. \ Phys. \ \textbf{84}, 839 (2012).

\bibitem{MDK94} C.~Mak, C.~Daly, and J.~Krim,
Thin \ Solid\ Films \textbf{253}, 190 (1994).

\bibitem{DAK98} A.~Dayo, W.~Alnasrallah, and J.~Krim,
Phys. \ Rev. \ Lett. \textbf{80}, 1690 (1998).

\bibitem{MK06} M.~Highland and J.~Krim,
Phys. \ Rev. \ Lett. \textbf{96}, 226107 (2006).

\bibitem{PBFMBMV10} M.~Pierno, L.~Bruschi, G.~Fois, G.~Mistura, C.~Boragno, F.B. de Mongeot, and U.Valbusa,
Phys. \ Rev. \ Lett. \textbf{105}, 016102 (2010).

\bibitem{KGGMRM94} M.~Kisiel, E.~Gnecco, U.~Gysin, L.~Marot, S.~Rast, and E.~Meyer,
Nature \ Mater. \textbf{10}, 119 (2011).

\bibitem{WYKHBW12} B.~Wolter, Y.~Yoshida, A.~Kubetzka, S.-W.~Hla, K.~von~Bergmann, and R.~Wiesendanger, 
Phys.\ Rev.\ Lett.\ \textbf{109}, 116102 (2012).

\bibitem{CWLSJ16} X.~Cai, J.~Wang, J.~Li, Q.~Sun, and Y.~Jia,
Trib.\ Inter.\ \textbf{95}, 419 (2016).

\bibitem{LG18} Y.~Li and W.~Guo, 
Phys.\ Rev.\ B.\ \textbf{97}, 104302 (2018).

\bibitem{KHW08} D.~Kadau, A.~Hucht, and D.~E.~Wolf, 
Phys.\ Rev.\ Lett.\ \textbf{101}, 137205 (2008).

\bibitem{Hucht09} A.~Hucht,
Phys.\ Rev.\ E.\ \textbf{80}, 061138 (2009).

\bibitem{AHW12} S.~Angst, A.~Hucht, and D.~E.~Wolf, 
Phys.\ Rev.\ E.\ \textbf{85}, 051120 (2012).

\bibitem{HA12} A.~Hucht and S.~Angst,  
Europhys.\ Lett. \textbf{100}, 20003 (2012).

\bibitem{IPT11} F.~Igl\'{o}i, M.~Pleimling, and L.~Turban, 
Phys.\ Rev.\ E.\ \textbf{83}, 041110 (2011).

\bibitem{Hilhorst11} H.~J.~Hilhorst,
J.\ Stat.\ Mech.\ P04009 (2011).

\bibitem{LP16} L.~Li and M.~Pleimling, 
Phys.\ Rev.\ E.\ \textbf{93}, 042122 (2016).

\bibitem{Sugimoto19} K.~Sugimoto,
Phys.\ Rev.\ E \textbf{99}, 052103 (2019).

\bibitem{FWN08} C.~Fusco, D.~E.~Wolf, and U.~Nowak,
Phys.\ Rev.\ B. \ \textbf{77}, 174426 (2008).

\bibitem{DD10} V.~D\'{e}mery and D.~S.~Dean,
Phys.\ Rev.\ L. \ \textbf{104}, 080601 (2010).

\bibitem{MBWN09} M.~P.~Magiera, L.~Brendel, D.~E.~Wolf, and U.~Nowak,
Europhys.\ Lett. \textbf{87}, 26002 (2009).

\bibitem{MBWN11} M.~P.~Magiera, L.~Brendel, D.~E.~Wolf, and U.~Nowak,
Europhys.\ Lett. \textbf{95}, 17010 (2011).

\bibitem{MAHW11} M.~P.~Magiera, S.~Angst, and A.~Hucht, and D.~E.~Wolf, 
Phys.\ Rev.\ B.\ \textbf{84}, 212301 (2011).




\bibitem{Ruina83} A.~Ruina, 
J. \ Geophys. \ Res. \textbf{88}, 10359 (1983)

\bibitem{Dieterich87}  J.~H.~Dieterich, 
Tectonophysics \textbf{144}, 127 (1987)

\bibitem{DK94} J.~H.~Dieterich, and B.~D.~Kilgore,
Pure \ and \ Appl. \ Geophys. \textbf{143}, 283 (1994)

\bibitem{HBPCC94} F.~Heslot, T.~Baumberger, B.~Perrin, B.~Caroli, and C.~Caroli,
Phys.\ Rev.\ E.\ \textbf{49}, 4973 (1994).

\bibitem{Scholz98} C.~H.~Scholz,  
Nature \textbf{391}, 37 (1998).

\bibitem{JMBSGKLJ16} V.~Jeudy, A.~Mougin, S.~Bustingorry, W.~Savero~Torres, J.~Gorchon, A.~B.~Kolton, A.~Lema\^{i}tre, and J.-P.~Jamet,
Phys.\ Rev.\ L.\ \textbf{117}, 057201 (2016).

\bibitem{DSKBJ17} R.~Diaz~Pardo, W.~Savero~Torres, A.~B.~Kolton, S.~Bustingorry, and V.~Jeudy,
Phys.\ Rev.\ B.\ \textbf{95}, 184434 (2017).

\end{thebibliography}
\end{document}